\documentstyle{article}

\title{{\bf Ac transport studies in polymers by a resistor network and transfer 
matrix approaches: application to polyaniline.
 }} 

\author{H.N. Nagashima\thanks{haroldon@ifsc.sc.usp.br}, 
R. N. Onody\thanks{onody@ifsc.sc.usp.br} 
and R. M. Faria\thanks{faria@ifsc.sc.usp.br} \\ \\ 
\small {\em Instituto de F\'{\i}sica de S\~{a}o Carlos} \\
\small {\em Universidade de S\~{a}o Paulo - Caixa Postal 369} \\
\small {\em 13560-970 - S\~{a}o Carlos, S\~{a}o Paulo, Brasil.}}
\date{}
\begin{document}
\maketitle
\normalsize
\baselineskip=24pt
\begin{abstract}

A statistical model of resistor network is proposed to describe a polymer
structure and to simulate the real and imaginary components of its ac 
resistivity. It takes into account the polydispersiveness of the material 
as well as intrachain and interchain charge transport processes. By the 
application of a transfer matrix technique, it reproduces ac resistivity 
measurements carried out with polyaniline films in different doping degrees 
and at different temperatures. Our results indicate that interchain processes 
govern the resistivity behavior in the low frequency region while, for higher 
frequencies, intrachain mechanisms are dominant.

\vspace{3cm}

\end{abstract}

\newpage

\section{Introduction}

The strong variation of electrical conductivity of conjugated polymer films
resulting from chemical doping, as seen in the insulator-to-metal transitions, 
was first observed in polyacetylene \cite{ckchi}.
Since then, a great amount of experimental and 
theoretical studies has been done to elucidate the mechanisms involved in the
electronic transport phenomena. 
Initial efforts were dedicated to understand the conduction mechanism along
a single chain, with the SSH model \cite{ssh}, a theoretical
one-electron framework based on the H\"uckel approach, which most 
successfully described the one-dimensional trans-polyacetylene. 
Doping mechanisms associated with chain defects brought into vogue the 
concepts of solitons and polarons as intermediate levels in the electronic 
structures of a single conjugated polymeric chain \cite{lu}. The number of 
such excitonic defects are directly related with the doping level: from weak 
to moderate doping the conduction mechanism is mainly due to electronic 
hopping, while 
extended states are reached by strongly doped samples. In macroscopic 
samples, for which extremely large conductivity changes can be observed 
(circa of 15 orders 
of magnitude), interchain processes must be taken into 
account owing to interactions brought about by the molecular packing. 
Theoretically, such interactions may be represented by the 
interchain exchange integral, which contribute significantly
to the final conductivity value.

The major difficulty in the study of bulk conductivity of polymeric 
materials lies in their highly disordered structure. The carrier transport 
should contain competing intrachain and interchain processes, and the 
challenge is not only to identify the contribution of each process, but also 
to quantify them. In this context, the alternating conductivity technique is 
an appropriate tool due to its spectroscopic character.

Among other theoretical treatments,
statistical models have also been applied to describe 
polymerization processes \cite{lia}, as well as, to study electrical 
properties
of polymers \cite{ba-sha,andra}. These models are essentially based on 
percolation concepts and have the advantage of simultaneously treating 
the phenomena involved on both a
microscopic and a macroscopic scale. Baughman and Shacklette proposed a 
random resistor network model which successfully explained the conductivity 
dependence on the conjugated length \cite{ba-sha}. 
Andrade et al \cite{andra} investigated the influence of percolation 
disorder on the electrical behavior in the dc regime. The 
distinction between intrachain and interchain mechanisms, both equally 
important to bulk conductivity, nevertheless requires a more detailed 
analysis.

In this paper we investigate the conductivity properties of weakly doped 
polyaniline in the presence of an alternating electric field, from both the 
theoretical and experimental perspectives. Ac conductivity measurements 
were carried out in polyaniline (PANI) films at different temperatures and 
doping degrees.
The system is theoretically described  by a percolation model where the 
basic ingredients responsible for the conductivity are incorporated. 
The model also takes into account the polydispersiveness of the material, 
the interchain and the intrachain charge transport. The real and imaginary 
parts of 
the resistivity were determined using a transfer matrix technique. At low 
frequencies, interchain processes are more important and dominate the 
transport 
mechanism. On the other hand, at high frequencies charge transport should
be restricted along the polymer chains, so intrachain 
processes should be dominant. Both regimes are well described by our model, 
which also reproduces the experimental results in a remarkable way.

\section{Theoretical Model and Numerical Simulations}

The theoretical model introduced in this paper allow us to focus two 
important aspects of the material: its geometrical structure and its 
electrical conductivity. 

First, we build a strip by juxtaposing $N$ square cells,
each cell with $L$ x $L$ sites (see Fig.1). 
Here, cells are used to conciliate the polymer size distribution, obtained
experimentally by the Gel Permeation Chromatography method,
with the transfer matrix technique developed by Derrida {\it et al.} 
\cite{derri}. The use of cells was applied here to calculate the electrical 
conductivity {\em without} excceeding the capacity of the computer's memory. 
In its original formulation 
\cite{derri}, the 
tranfer matrix acts on a random and {\em ramified} structure of resistors.
Here, however, only a non ramified structure is possible. Therefore, we 
construct a cell which contains only linear structures with a Gaussian size 
distribution and fixed density. The information of one cell is 
stored and transferred, by the matrix method, to the next cell.

An empty lattice site $i$ of the cell is randomly occupied, representing the 
presence of a monomer. We shall call this site the {\em growing tip}.
The chain size $l$ is taken from a {\em Gaussian} distribution probability 
centered in $l_{0}$ with dispersion $\Delta l$. An {\em empty} site $j$, 
nearest 
neighbor of the growing tip $i$, is randomly occupied and a resistance 
$R_{c}$ is attributed to the bond connecting these two monomers (sites).
The site $j$ becomes now the new growing tip. The process is repeated until 
the chosen polymer size is reached. If the polymer cannot reach the specified 
length, it is discarded. Clearly, a self-avoiding scheme is used to mimic the 
excluded volume interaction. 

After the chain has reached the selected size, 
a search in all of its monomers is performed looking for a pair of nearest 
neighbor monomers (which can belong or not to the same chain), not connected 
by $R_{c}$.
If such a pair is found then these monomers are connected,
with probability $p_{d}$, by a $RC$ circuit.
This $RC$ circuit 
is a parallel association where $C$ is the capacitance and $R_{i}$ is the
interchain resistance . The idea is to incorporate  
induced charges and hopping (or tunneling) mechanisms. As we shall see, 
$p_{d}$ is related to the doping level and we call it the doping parameter. 

Repetition of the procedure described above
built new polymer chains inside the cell until some concentration $k$ 
of the occupied lattice bonds is reached. Let $ N_{ob}$ 
be the number of occupied lattice bonds, i. e., $N_{ob}$ = number of  
resistances $R_{c}$ + number of impedances $Z$ (the equivalent impedance of
the circuit $RC$) and $N_{eb}$ the
number of empty lattice bonds (corresponding to infinite resistances) then
$k$ = $N_{ob}$/($N_{ob} + N_{eb}$). If $k=0.5$ the system is at the critical
bond percolation threshold of the square lattice.
A typical cell is depicted in Fig.1. It is considered that a bias voltage is
applied on both sides, being $L$ the distance between the electrodes.

To determine the longitudinal resistivity $\rho_{//}$ of the cell, which is the 
resistivity in the direction of the current flow, we follow the tranfer matrix 
method \cite{derri}. 
Compared to the two other main approaches - resolution of Kirchoff's equations 
and node elimination, this technique has many advantages.
It consumes less CPU time, it avoids troubles with dangling bonds or isolated 
clusters, and the conductivity is calculated exactly. Of course, care should be
taken when passing from one cell to another: horizontal and 
vertical bonds {\em on the boundary} must be stored in order to correctly 
construct
the next cell. We used around $ N \sim 100,000 $ cells in our simulations.

Simulated curves $\rho_{//}$ versus $f$ (the frequency of an ac electric field)
are presented in Fig.2. 
They exhibit two plateaus, i. e., two regions in which the resistivity is 
frequency-independent: one at low frequencies and the other at high 
frequencies. 
Between these two plateaus the resistivity decays with the frequency, obeying 
approximately the relation $\rho_{//} \propto f^{-\alpha}$ with $\alpha = 1.9$. 
It does not agree with the measured value $\alpha \simeq 1.2 $, which is the 
slope of the experimental curves of Fig. 3.
This discrepancy, however, disappears when we consider, as we shall see later, 
the interchain resistance $R_{i}$ dependence with the frequency $f$.
Figures 2a, 2b and 2c correspond to simulations performed at concentration 
$k=0.5$ and they show, respectively, the influence of $p_{d}$, $R_{c}$ and 
$R_{i}$ in the $\rho_{//}(f)$ curves. $R_{i}$ acts in the low frequency 
plateau, 
$R_{c}$ in the high frequency and $p_{d}$ causes the whole curve to shift up 
and down. The interpretation of these parameters will be discussed below, when
the simulated results and the experimental curves carried out in PANI films
are compared. Figure 2d was obtained at fixed values of 
$p_{d}$, $R_{c}$ and $R_{i}$ but for two different concentrations $k=0.5$ 
and $k=0.6$. Increasing $k$ simply displaces the whole curve
of resistivity downward in a frequency independent way. This is expected since,
to increase the concentration $k$ is equivalent to substitute a number of empty
bonds (infinite resistances) by occupied bonds (finite resistances). Note
that the effects of increasing $k$ or $p_{d}$ are very similar - both are 
frequency independent and decrease the resistivity. 
As the 
concentration $k$ does not essentially alter the form of the 
curves, in the rest of this paper we restrict the simulations to be performed 
at $k=0.5$.

\section{Experimental Resistivity}

Responses to alternating electrical field applied in doped conjugated polymers 
have been an important subject of research when charge transport phenomena are 
involved. 
Some ac results, conductivity versus frequency, have already been reported in 
the literature, with polyacetylene \cite{epstein,ito}, poly(3-methylthiophene) 
\cite{reh} and poly(o-methoxyaniline) \cite{faria}. Here we present some 
results 
carried out with polyaniline films for different doping degrees and different 
temperatures. Polyaniline is a common name for a family of polymers consisting 
of a sequence of oxydized [-(C6H4)-N=(C6H4)=N-] and reduced 
[-(C6H4)-(NH)-(C6H4)-(NH)-] units. Three distinct forms 
have been identified and the emeraldine form, which contains an equal number of 
oxydized and reduced segments, has been the most frequently studied. 
Polyaniline was 
synthesized and processed as described elsewere \cite{macdiarmid}. 
Flexible and free-standing films were obtained. Upon protonation, plunging the 
emeraldine sample into aqueous HCl solutions of different molarities, 
its resistivity changes from the characteristic of poorly conducting 
semiconductors 
($ \rho \sim 10^{10} \; \Omega $.cm ) to metals ($ \rho \sim 10^{-1}-10^{-3} \; 
\Omega $.cm ). 
Circular,$ 12 \; \mu$m in thick samples, had central gold 
electrodes
of a diameter equal to $ 0.5 $ cm evaporated on both sides. Impedance 
measurements were carried out using a controlled Frequency Response 
Analyser coupled to a Potentiostat which operates in the $10^{-1}$ to $10^{6}$ 
Hz frequency range.

Figure 3 shows measurements of real and imaginary components of ac 
resistivity of PANI films under conditions: a) doped in HCl solution of
$0.1 M$ at $300 K$ (measurement denoted by $M_{1}$); b) doped in HCl of
$0.01 M$ at $300 K$ ($M_{2}$); and c) doped in HCl of $0.1 M$ at $200 K$ 
($M_{3}$).
It should be that the real and imaginary parts of the longitudinal resistivity 
$ \rho_{//} $ are proportional to these measured resistivities \cite{tareev}.
The sample used in the curves of Fig. 3 (b) is less doped than that of Fig. 3 
(a), while the sample in Fig. 3 (c) is equally doped but it is obtained at a 
lower temperature.

\section{Results and Discussion}

The curves of Fig.3 are faithfully reproduced by our model by fixing 
$R_{c}$ and $p_{d}$ 
values and having a frequency-dependent interchain resistance $R_{i}$ as 
shown in the insets of Fig. 4a. Fig. 4b ($M_{2}$ measurement) is obtained
with $R_{c} = 2.0 \omega$ and $p_{d} = 0.18$. Finally for $M_{3}$, 
$R_{c} = 0.5 \omega$ and $p_{d} = 0.21$ - the same values as in $M_{1}$ the
differences being on account of the hopping resistance $R_{i}$. At low
frequencies $R_{i}$ of $M_{3}$ is ten times bigger than that of $M_{1}$ (see
insets of Fig. 4a and 4c). Then from these results we conclude that: $i)$
the increase in the doping can either be simulated by diminishing $R_{c}$ or
increasing $p_{d}$ and $ii)$ decreasing the temperature the sample resistance
increases, as expected.
The resistance $R_{c}$ represents the conduction mechanism 
along a single chain. It should depend only slightly on the temperature (see 
for 
example the SSH model \cite{ssh}), but it is sensitive to the doping degree. 
The doping 
parameter $p_{d}$ changes with the doping degree (as can be seen in curves 
4(a) and 4(b)), but it is frequency independent.
In a highly disordered bulk structure, on the other hand, 
interchain charge transfer processes 
of electronic carriers can be represented by a distribution of energy barriers 
$W$ of interchain hopping (or phonon-assisted tunneling). Since $R_{i}$ 
represents, in our model, the interchain processes, its dependence on the 
frequency is explained in terms of this energy barrier distribution.
A given waiting time $\tau$ is 
associated with an energy barrier $W$ by $ \tau \sim  exp(W/kT) $.
In the low frequency region, where $ f < \tau^{-1} $, electron motion is 
hampered mainly by the high energy barriers (high $W$), which represent 
interchain obstacles. For $ f > \tau^{-1} $, on the other hand, the electronic 
carriers become localized in small regions of low energy barriers, being 
therefore, more mobiles. 
This description agrees qualitatively with the frequency-variation of $R_{i}$ 
shown in the insets of Fig.4. To have a more formal insight on the subject,
let us consider the random free-energy barrier model.

This model is an elegant way to approach ac conductivity in polymeric
disordered systems. In this formalism all information about disorder is 
contained in the hopping time distribution function $\psi (t)$. According to
J. C. Dyre \cite{dyre} $\psi (t)$ is determined by hopping over random 
distributed energy.
He used the continuous time random walk and the effective medium approximation 
to solve the model and to derive a simple expression for the complex 
conductivity $\sigma$ as a function of the frequency $f$

\begin{center}
$\sigma (f) = C [ -i f + \frac{i f \ln (\frac{\gamma_{max}}{\gamma_{min}})}
{\ln (\frac{1+i f / \gamma_{min}}{1+i f / \gamma_{max}})}] 
$
\end{center}
where $C$ is a constant depending among other parameters on the density of
carriers, $\gamma_{max}$ and $\gamma_{min}$ are two parameters related to the 
jump frequency. 

In our model, the $R_{i}$ dependence with $f$ was determined by 
adjusting the real and the imaginary parts of 
the simulated resistivity, for each frequency, to the experimental values. 
This give us
an infinity of fitting parameters. But now we can compare our results
with those predicted by Dyre's solution.
The advantage is that Dyre's hopping resistance has only three fitting 
parameters for all frequencies. Figure 5 shows that the adjusted $R_{i}$
is in reasonable agreement with Dyre's theory.
   
Extrapolating the dependence of $R_{i}$ with $f$ for {\em high} frequencies 
(unattainable in our equipment), as shown in the insets of Fig. 4, our 
simulations anticipate the emergence of a second plateau for the real component 
of the resistivity. This prediction is compatible with other results found for 
a polymeric blend \cite{yoon}.

\section{Summary and Conclusions}

In conclusion, we introduce a model which carefully takes into account the 
geometric aspects of the real polymeric structure. It considers the polymer 
chains 
being constructed from a Gaussian size distribution and in the presence of the 
excluded volume interaction. It incorporates, in a simple way, the existence of 
both intrachain and interchain charge transport mechanisms. Simulations
based on a tranfer matrix technique and performed at the critical percolation 
threshold, reproduces the measured resistivity in PANI films in a large range 
of frequencies for different doping and temperature conditions. 
Estimatives of the critical conductivity exponent using a finite size scaling 
approach will be the object of a forthcoming paper.

\section{Acknowledgments}

We acknowledge CNPq (Conselho Nacional de Desenvolvimento
Cient\'{\i}fico e Tecnol\'ogico) and FAPESP ( Funda\c c\~ao de Amparo 
a Pesquisa do Estado de S\~ao Paulo ) for the financial support.

\begin{center}

{\bf Figure Captions}

\end{center}

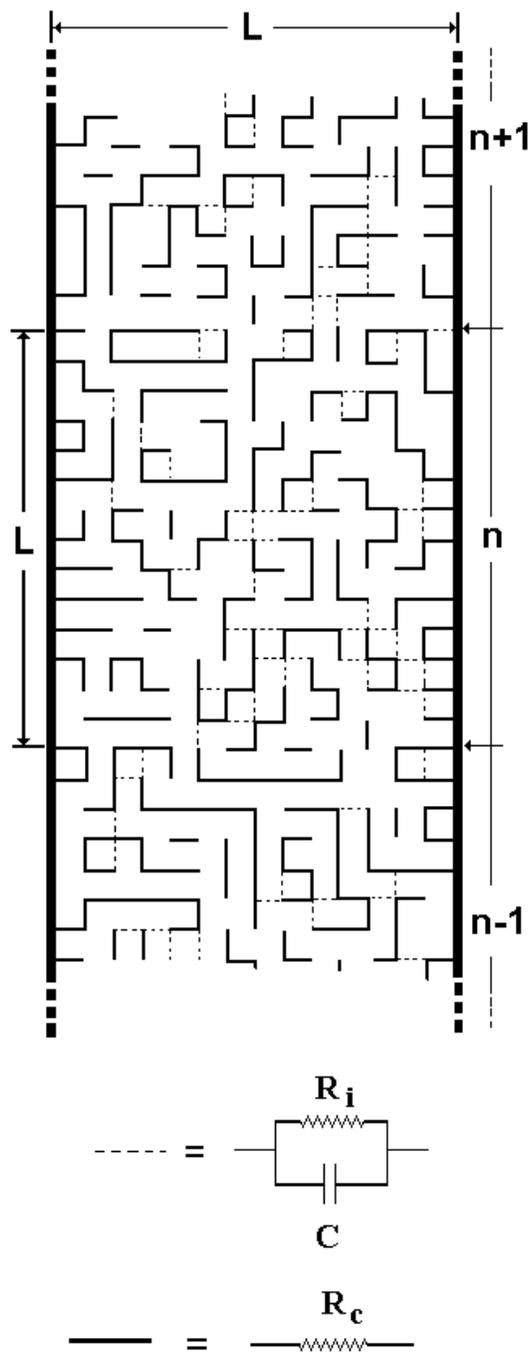
Figure 1. The nth cell of a typical strip. Here $L=15$ and we use a gaussian 
size distribution centered in $l_{0}=6$ and width $\Delta l=6$. Continuous 
lines represent the polymer chains ($R_{c}$ resistors) and broken lines the 
interchain bridges ($RC$ circuits). The two thick lines are the electrodes. 

\vspace{1.5cm}

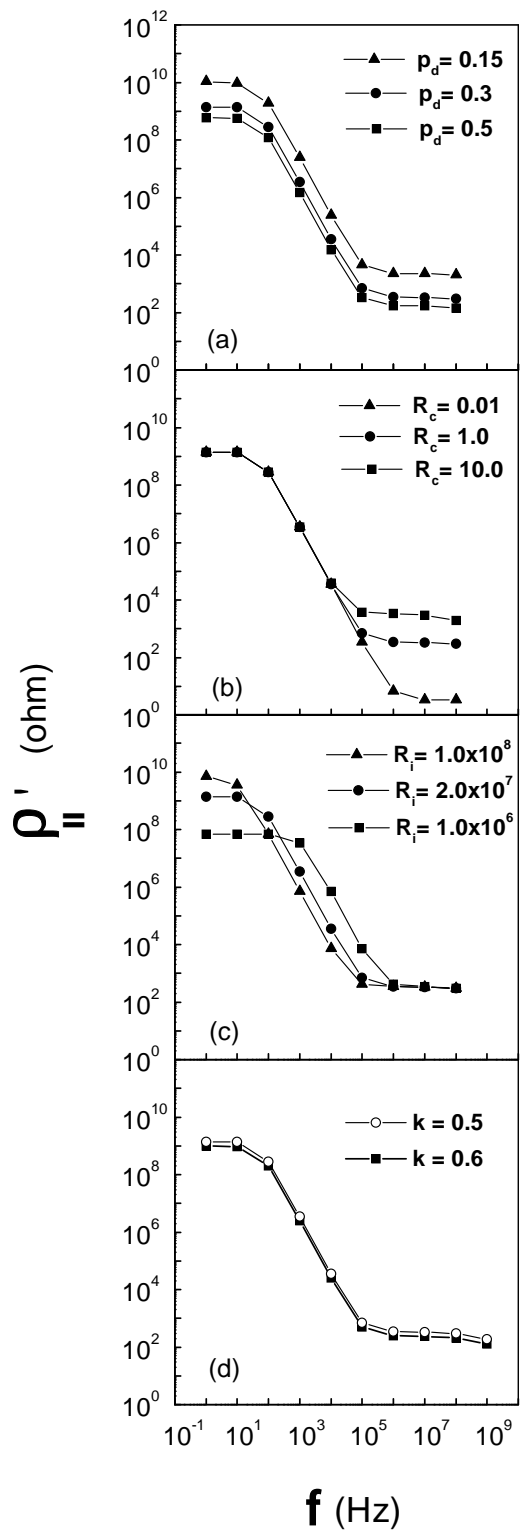
Figure 2. The longitudinal resistivity $ \rho_{//} $ as a function of the 
frequency $f$ for several values of: (a) doping parameter $p_{d}$; (b) 
intrachain resistance $R_{c}$; (c) interchain resistance $R_{i}$ and fixed 
capacitance $ C = 1$ $ nF $; (d) concentration $k$.

\vspace{1.5cm}

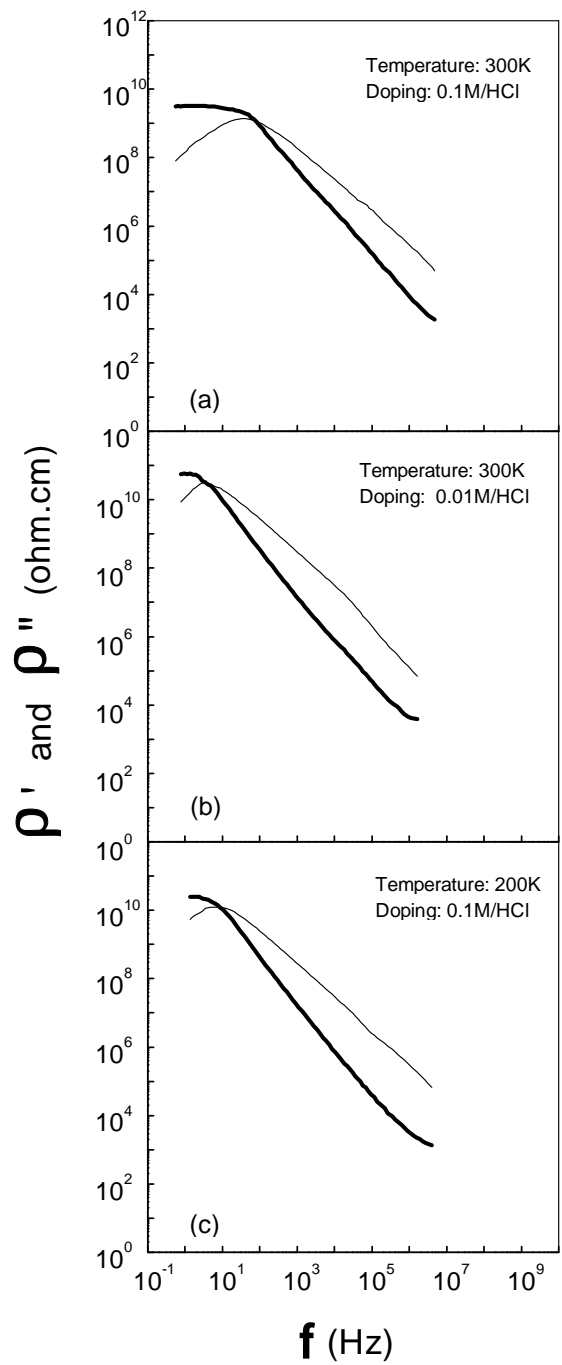
Figure 3. Dependence of the real and imaginary parts ($ \rho ' $ and $ 
\rho '' $ which are represented by thick and thin lines, respectively) of the 
measured polyaniline resistivity with the frequency $f$ for different 
temperatures and doping degrees.

\vspace{1.5cm}

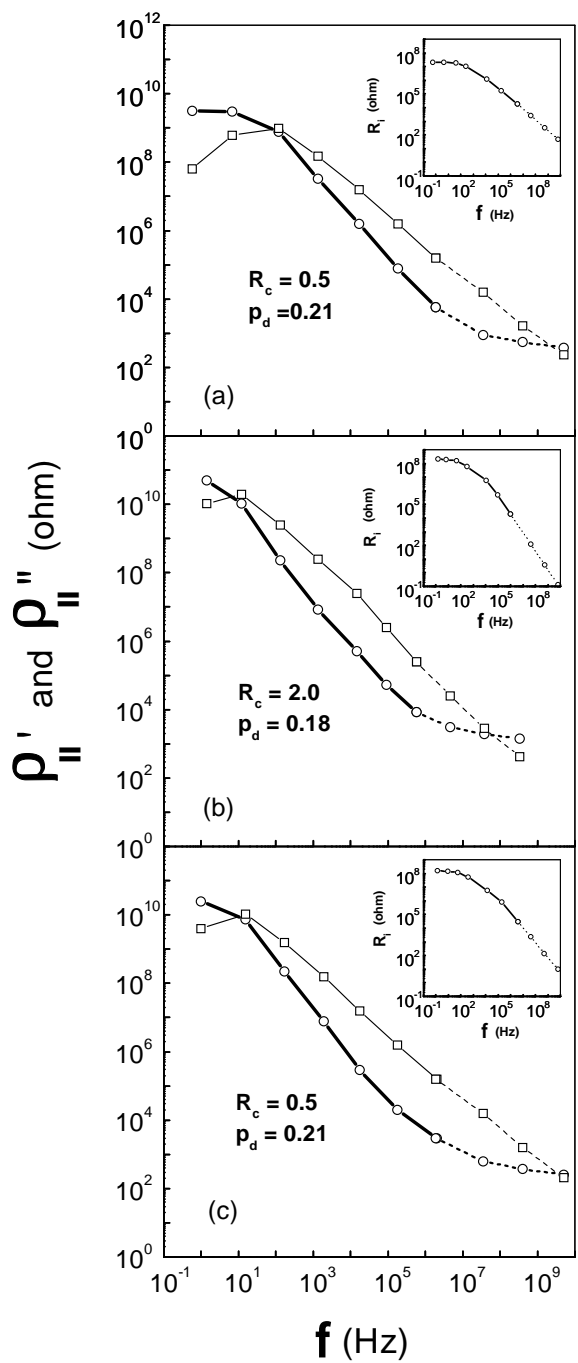
Figure 4. Plots of the real and imaginary parts ($ \rho_{//} ' $ and $ 
\rho_{//} '' $) of the simulated longitudinal resistivity (thick and thin 
lines, respectively) as a function of the frequency $f$. To reproduce the 
experimental curves, the interchain resistance $R_{i}$ should depend on the 
frequency as shown in the insets. The broken lines corresponds to extrapolated 
values as explained in the text.

\vspace{1.5cm}

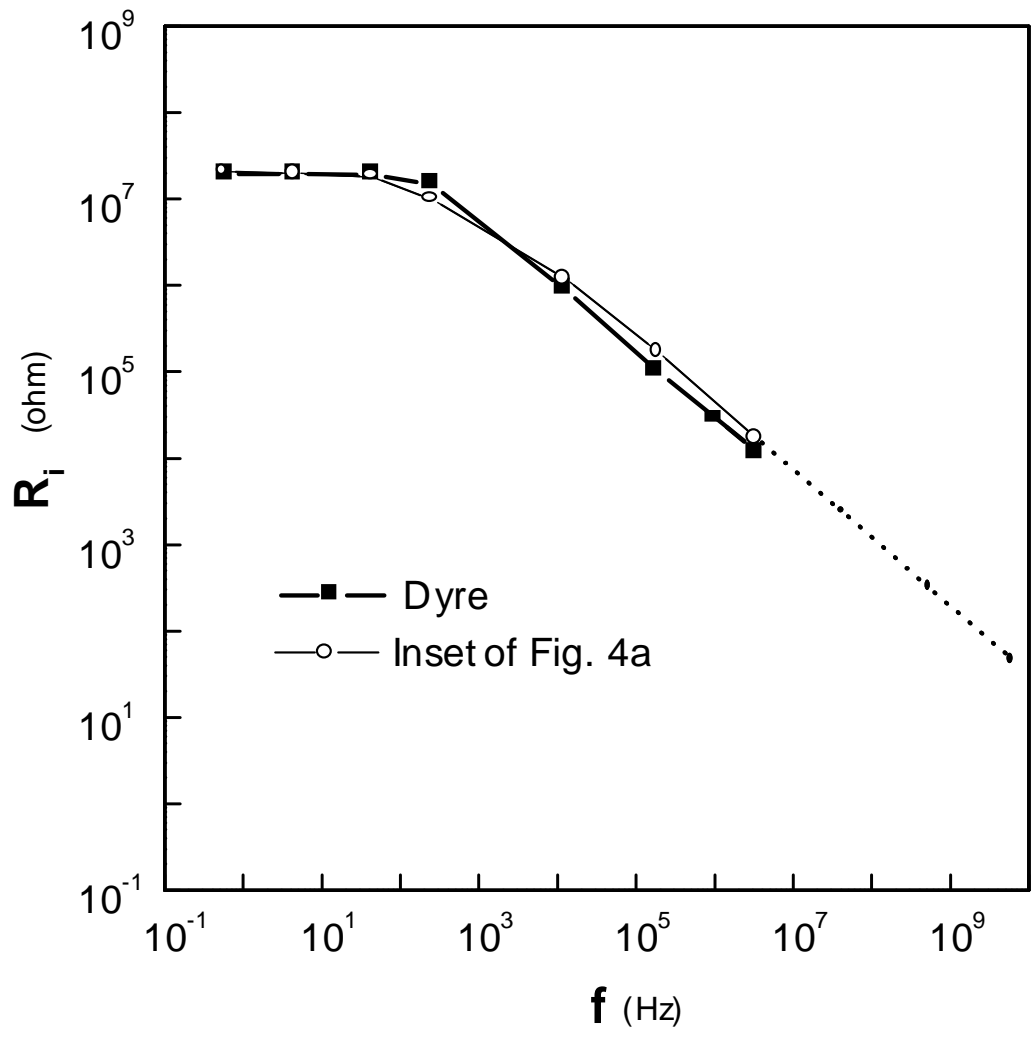
Figure 5. Dyre's hopping resistance ($Re(\frac{1}{\sigma})$) and the
adjusted $R_{i}$ (inset of Fig. 4a) versus the frequency $f$. The values
of the parameters $C$, $\gamma_{min}$ and $\gamma_{max}$ are 
$ 1.0 \: 10^{10}$, $200$ and $1.0 \: 10^{9}$.

\end{document}